\documentclass[12pt,a4paper]{article}

\usepackage{amsmath,amsfonts}
\usepackage{amssymb}
\usepackage{cite}
\usepackage{epsfig,psfrag}
\usepackage[usenames,dvips]{color}

\usepackage[margin=3.5cm]{geometry}

\makeatletter
\@addtoreset{equation}{section}
\makeatother

\newcommand{\be}{\begin{equation}}
\newcommand{\ee}{\end{equation}}
\newcommand{\bea}{\begin{eqnarray}}
\newcommand{\eea}{\end{eqnarray}}

\newcommand{\h}{\mathfrak h}

\begin{document}
\thispagestyle{empty}

\begin{center}
\hfill CPHT-RR096.1110\\
\hfill UAB-FT-684

\begin{center}

\vspace{1.7cm}

{\LARGE\bf 
Warped Electroweak Breaking Without Custodial Symmetry
}

\end{center}

\vspace{1.4cm}

{\bf Joan A. Cabrer$^{\,a}$, Gero von Gersdorff$^{\;b}$ 
and Mariano Quir\'os$^{\,a,\,c}$}\\

\vspace{1.2cm}
${}^a\!\!$ {\em {Institut de F\'isica d'Altes Energies (IFAE), Universitat Aut{\`o}noma de Barcelona\\
08193 Bellaterra, Barcelona, Spain}}

\vspace{.1cm}

${}^b\!\!$ {\em {Centre de Physique Th\'eorique, \'Ecole Polytechnique and CNRS\\
F-91128 Palaiseau, France}}

\vspace{.1cm}

${}^c\!\!$ {\em {Instituci\'o Catalana de Recerca i Estudis  
Avan\c{c}ats (ICREA)}}

\end{center}

\vspace{0.8cm}

\centerline{\bf Abstract}
\vspace{2 mm}
\begin{quote}\small
We propose an alternative to the introduction of an extra gauge (custodial) symmetry to suppress the contribution of KK modes to the $T$~parameter in warped theories of electroweak breaking. The mechanism is based on a general class of warped 5D metrics and a Higgs propagating in the bulk. The metrics are nearly AdS in the UV region but depart from AdS in the IR region, towards where KK fluctuations are mainly localized, and have a singularity outside the slice between the UV and IR branes. This gravitational background is generated by a bulk stabilizing scalar field which triggers a natural solution to the hierarchy problem. Depending on the model parameters, gauge-boson KK modes can be consistent with present bounds on EWPT for $m_{\rm KK}\gtrsim 1$ TeV at 95\% CL. The model contains a light Higgs mode which unitarizes the four-dimensional theory. The reduction in the precision observables can be traced back to a large wave function renormalization for this mode.

\end{quote}

\newpage
%%%%%%%%%%%%%%%%%%%%%%%%%%%%%%%%%%
{\it 1. \underline{Introduction}} 

Warped models in five dimensions were proposed by Randall and Sundrum (RS)~\cite{RS1} as an elegant way of solving the Planck/weak hierarchy problem. In the RS setup, we live in a slice of AdS$_5$ with two flat branes in the ultraviolet (UV) and infrared (IR) regions whose four-dimensional theory is Poincar\'e invariant. If the Higgs field is localized on (or towards) the IR brane,  in the 4D theory its Planckian mass is warped down to the TeV scale, and this is how the large hierarchy is generated. 

When gauge bosons propagate in the 5D bulk their Kaluza-Klein (KK) excitations can contribute to the Standard Model electroweak precision observables --in particular the $T$~parameter~\cite{Huber:2000fh,Davoudiasl:2009cd}-- and their masses and couplings have to be contrasted with the Standard Model electroweak precision tests (EWPT). Since KK modes decouple when they are heavy, EWPT translate into lower bounds on their masses. If these bounds are much larger than LHC scales, they make the theories phenomenologically unappealing and create a ``little hierarchy'' problem, which translates into some amount of fine-tuning to stabilize weak masses. In particular, in order to avoid large volume-enhanced contributions to the $T$~parameter it was proposed to enlarge the gauge symmetry in the bulk by adding the $SU(2)_R\times U(1)_{B-L}$ gauge group~\cite{Agashe:2003zs}. In this case one has to worry only for large contributions to the $S$~parameter, which yield bounds on KK masses of $\mathcal O(3)$ TeV. Moreover, the presence of extra matter, in particular the $SU(2)_R$-symmetric partner of the RH top quark, generate large anomalous (volume enhanced) contributions to the $Z\bar bb$ vertex which should be controled by introducing a discrete left-right symmetry~\cite{Agashe:2006at}. 

The presence of large IR brane kinetic terms was proposed in Ref.~\cite{Davoudiasl:2002ua} as a way of reducing the $T$~parameter. Generically they include bare contributions, which encode physics above the cutoff scale, as well as radiative contributions that are calculable within the effective 5D theory.  However on the IR brane radiative effects are small since the local cutoff is around the TeV scale, there is no room for a large logarithmic enhancement and a large IR brane kinetic term would have to arise from the unknown UV physics.
Let us remark though that such unknown UV physics could just as well directly modify the $T$ parameter, by contributing to an IR brane localized operator of the type $|H^\dagger D_\mu H|^2$. In order to keep calculability we will assume
such degrees of freedom to be absent.

In this letter we will propose a simple alternative to the introduction of an extra custodial symmetry (or large IR brane kinetic terms) based on generalized metrics and a bulk Higgs field, to keep the $T$~parameter under control, which should allow us to construct a pure 5D Standard Model. Although there are negative results in the literature based on general 5D metrics when the Higgs is localized on the IR brane~\cite{Delgado:2007ne,Archer:2010hh}, we will circumvent them by assuming a Higgs propagating in the bulk of the fifth dimension in the background of a singular metric (asymptotically AdS near the UV brane) with a singularity outside the slice between the branes, but close enough to the IR brane. In fact, as we will see, the $T$~parameter will be suppressed by the combined effect of the non-localized bulk Higgs as well as the vicinity of the singularity to the IR brane. This effect can be traced back to a large wave function renormalization for a light Higgs mode. 
This alternative to the custodial symmetry is purely based on a modification of the 5D gravitational background, which requires a bulk propagating scalar field playing the role of the Goldberger-Wise field~\cite{GW} in RS theories, and does not interfere with the electroweak physics. This kind of metrics have been widely studied in the literature in the past~\cite{Gubser:2000nd} and more recently they constitute the background of the so-called soft-wall models~\cite{AdS/QCD, Falkowski1,Falkowski2,Batell2,Delgado:2009xb,MertAybat:2009mk,Cabrer:2009we,Atkins:2010cc,vonGersdorff:2010ht}. Let us mention that a reduction in the S parameter in a particular class of soft-wall models with custodial symmetry was already pointed out in Ref.~\cite{Falkowski1}.

\

{\it 2. \underline{General results}} 

We will now consider the Standard Model (SM) propagating in a 5D space with an arbitrary metric $A(y)$ such that
\be
ds^2=e^{-2 A(y)}\eta_{\mu\nu}dx^\mu dx^\nu+dy^2\,,
\ee
in proper coordinates and two branes localized at $y=0$ and $y=y_1$, at the edges of a finite $S^1/\mathbb Z_2$-interval. We define the 5D $SU(2)_L\times U(1)_Y$ gauge bosons as $W^i_M(x,y)$, $B_M(x,y)$ [or in the weak basis $A_M^\gamma(x,y)$, $Z_M(x,y)$ and $W^{\pm}_M(x,y)$], where $i=1,2,3$ and $M=\mu,5$, and the SM Higgs as
\be
H(x,y)=\frac{1}{\sqrt 2}e^{i \chi(x,y)} \left(\begin{array}{c}0\\h(y)+\xi(x,y)
\end{array}\right)
\ ,
\label{Higgs}
\ee
where the matrix $\chi(x,y)$ contains the three 5D SM Goldstone bosons. The Higgs background $h(y)$ as well as the metric $A(y)$ will be for the moment arbitrary functions which will be specified later on.

We will consider the 5D action (in units of the 5D Planck scale $M$) for the Higgs field $H$ and other possible scalar fields of the theory, generically denoted as $\phi$:
\begin{eqnarray}
S_5&=&\int d^4x dy\sqrt{-g}\left(-\frac{1}{4} \vec W^{2}_{MN}-\frac{1}{4}B_{MN}^2-|D_M H|^2-\frac12(D_M \phi)^2
-V(H,\phi)
\right)\nonumber\\
&-&\sum_{\alpha}\int d^4x dy \sqrt{-g}\,(-1)^\alpha\,2\,\lambda^\alpha(H,\phi)\delta(y-y_\alpha)
,
\label{5Daction}
\end{eqnarray}
where $V$ is the 5D potential and $\lambda^\alpha$ ($\alpha=0,1$) the 4D brane potentials which depend on the scalar fields of the theory. For the sake of simplicity, we will assume all the fermions of the SM living on the UV brane.
One can then construct the 4D effective theory out of (\ref{5Daction}) by making the KK-mode expansion~\cite{CGQ3}
$
A_\mu(x,y)=a_\mu(x)\cdot f_A(y)/\sqrt{y_1}
$
where $A=A^\gamma,Z,W^{\pm}$ and the dot product denotes an expansion in modes. The functions $f_A$ satisfy the equations of motion (EOM) 
\be
m_{f_A}^2 f_A+(e^{-2A}f^{\prime}_A)'-M_A^2 f_A=0
\ ,
\label{eq.f}
\ee
where the functions $f_A(y)$ are normalized as
$
\int_0^{y_1}f_A^2(y)dy=y_1
$
and satisfy the boundary conditions
$
e^{-2A}\left. f^{\prime}_A\right|_{y=0,y_1}=0.
$
We have defined the 5D $y$-dependent gauge boson masses
\be
M_W(y)=\frac{g_5}{2} h(y)e^{-A(y)}
\ ,
\qquad M_Z(y)=\frac{1}{c_W} M_W(y)
\ ,
\qquad M_\gamma(y)
\equiv 0
\ .
\label{masass}
\ee
where $c_W={g_5}/{\sqrt{g_5^2+g_5'^2}}$, and $g_5$ and $g'_5$ are the 5D $SU(2)_L$ and $U(1)_Y$ couplings respectively.

Equations \eqref{eq.f} will in general not have analytic solutions. However we can find an approximation for the light mode of \eqref{eq.f} with mass $m_{f_A^0}\equiv m_A$ in the limit where the breaking is small ($m_A\ll m_{KK}$) and thus there is a zero mode with almost constant profile. Expanding around this limit we can write
%
%\be
$f_A(y)=1-\delta_A+\delta f_A(y)
\ ,
%\label{fAequal}
$
%\ee 
%
where
\begin{eqnarray}
 \delta f_A(y) &=& \int_0^y dy'\,e^{2A(y')} \int_0^{y'} dy''\left[ M_{A}^2(y'') - m_{A,0}^2 \right]   
\ ,
 \\
 \delta_A&=& \frac{1}{y_1}\int_0^{y_1}  dy\, \delta f_A(y)
\ ,
 \\
 m_{A,0}^2 &=&  \frac{1}{y_1}\int_0^{y_1} {M^2_{A}(y)} dy 
\ ,
 \label{eq:fzeromodeapprox}
 \end{eqnarray}
and where $m_{{A,0}}$ is a zeroth order approximation for the zero mode mass $m_A$.
Including the first order deviations from a constant profile we obtain for the light mode mass the expression  
\be
 m^2_{A} =m_{A,0}^2-\frac{1}{y_1}\int_0^{y_1}dy\, M^2_{A}(y)\left[\delta_A-\delta f_A(y)   \right]
 \ .
\label{masa1}
\ee

\ 

In the SM the three Lagrangian parameters $(g,g',v)$, or equivalently $(e,s_w,v)$ are normally traded for the well measured parameters $(\alpha,G_F,m_Z)$ in such a way that all other SM observables can be written as functions of them. In our 5D model we have to fix the physical $Z$ mass using Eq.~(\ref{masa1})
$m_Z = 91.1876\pm 0.0021~\mathrm{ GeV}$ which provides a relation between the model parameters. To compare with EWPT a convenient parametrization is using the $(S,T,U)$ variables in Ref.~\cite{Peskin:1991sw}. They can be given the general expressions~\cite{CGQ3}:
\begin{eqnarray}
\alpha T&=&s^2_W \frac{m_Z^2}{\rho^2} k^2y_1\,\int_0^{y_1}\left(1-\Omega(y)\right)^2e^{2A(y)-2A(y_1)}\label{T}
\ ,
\\
\alpha S&=&8 c^2_W s^2_W \frac{m_Z^2}{\rho^2}k^2y_1\,\int_0^{y_1} \left(1-\frac{y}{y_1}\right)\left(1-\Omega(y)\right)e^{2A(y)-2A(y_1)}\label{S}
\ ,
\\
\alpha U&=&\mathcal O(\delta_Z^2)\simeq 0\label{U}
\ ,
\end{eqnarray}
where
\be
\rho=ke^{-A(y_1)},\quad \Omega(y)=\frac{U(y)}{U(y_1)},\quad U(y)=\int_0^y h^2e^{-2A}\ ,
\label{omegafunction}
\ee
We have reexpressed the hierarchy $\exp[-A(y_1)]$ by the ratio of an IR scale $\rho$ and a UV scale $k$ which we will take to be of the order of the TeV and Planck scales respectively. The parameter $\rho$ is related to the gauge boson KK-mode mass $m_{KK}$ by $m_{KK}=F(A)\rho$ where $F$ is a function which depends on the metric $A$ and it is entirely determined from the solution to the EOM (\ref{eq.f}). The function $\Omega$ is monotonically increasing from $\Omega(0)=0$ to $\Omega(y_1)=1$. In the case of an IR brane localized Higgs it is actually a step function and in particular it vanishes identically in the bulk, $\Omega=0$. More generally, due to the presence of the warp factor, the integral will be dominated near the IR and one could approximate
\be
\Omega(y)\simeq 1-\frac{k(y_1-y)}{Z}\,,
\ee
where 
\be
Z=\frac{k}{\Omega'(y_1)}=k\int_0^{y_1}dy\frac{h^2(y)}{h^2(y_1)}e^{-2A(y)+2A(y_1)}\ .
\label{Xgeneric}
\ee
The integrals in Eqs.~(\ref{T})-(\ref{U}) will be approximated well whenever $Z$ is large enough. One finds
\begin{eqnarray}
\alpha T&=&s^2_W \frac{m_Z^2}{\rho^2} ky_1 \frac{1}{Z^2}
\,I ,
\label{Tapp}
\\
\alpha S&=&8 c^2_W s^2_W \frac{m_Z^2}{\rho^2}\frac{1}{Z}\,I
\,,
\label{Sapp}
\end{eqnarray}
with the dimensionless integral
\be
I=k\int _0^{y_1} \left[k(y_1-y)\right]^2e^{2A(y)-2A(y_1)} dy
\ .
\ee
Notice in particular the standard volume enhancement of the $T$~parameter.
This approximation is valid whenever $Z$ is a parametrically large number (but not necessarily as large as the volume $ky_1$). One can see that the $T$~parameter is suppressed with two powers of $Z$, while the  $S$ parameter is only suppressed with one power. As we will see below, in theories with a light Higgs mode the quantity $\sqrt{Z}$ can in fact be interpreted as a wave function renormalization in the effective Lagrangian of that mode. Note that the operators contributing to $T$ have four powers of the Higgs field (e.g.~$|H^\dagger D_\mu H|^2$), while the ones contributing to $S$ have only two powers ($e.g.~H^\dagger W_{\mu\nu}HB^{\mu\nu}$), thus nicely explaining the observed suppression in models with large values of $Z$. In case of pure RS with a bulk Higgs profile $h(y)\sim e^{aky}$ one can easily evaluate $Z^{-1}=2(a-1)$, and in the region of interest that solves the hierarchy problem, $a>2$, this is not a small number. Nevertheless, we will see below that there are theories which can have sizable $Z$ factors despite the fact that $a>2$, and hence display suppression of the precision observables. In the remainder of this Letter we will always use for numerical calculations the exact expressions, Eq.~(\ref{T}) and (\ref{S}).

The SM fit on the $(S,T)$ plane, assuming $U=0$, for a reference Higgs mass $m_H^{ref}=117$ GeV, provides~\cite{Nakamura:2010zzi}
\begin{equation}
T = 0.07\pm 0.08 ,~~~
S = 0.03\pm 0.09,
\label{fit}
\end{equation}
which should constrain any particular model.

For the Higgs fluctuations in (\ref{Higgs}) one can write from the action (\ref{5Daction}) an EOM similar to that of gauge bosons (\ref{eq.f}). In fact making the KK-mode expansion $\xi(x,y)= H(x)\cdot \xi(y)/\sqrt{y_1}$  the functions $\xi(y)$ satisfy the bulk equation
\be
\xi''(y)-4A'\xi'(y)-\frac{\partial^2 V}{\partial h^2}\xi(y)+m_\xi^2 e^{2A}\xi(y)=0
\ ,
\label{Higgsbulk}
\ee
as well as the boundary conditions (BC)
\be
\frac{\xi'(y_\alpha)}{\xi(y_\alpha)}=\left.\frac{\partial^2 \lambda^\alpha(h)}{\partial h^2}\right|_{y=y_\alpha}
\ .
\label{HiggsBC}
\ee
If we compare Eq.~(\ref{Higgsbulk}) with the bulk EOM for the Higgs background~\cite{CGQ3} for a quadratic bulk Higgs potential
\be
h''(y)-4A'h'(y)-\frac{\partial V}{\partial h}=0
\ ,
\label{bckbulk}
\ee
with BC
\be
\frac{h'(y_\alpha)}{h(y_\alpha)}=\left.\frac{\partial\lambda^\alpha(h)}{\partial h}\right|_{y=y_\alpha}
\ ,
\label{bckBC}
\ee
we see that the Higgs wave function for $m_H\equiv m_\xi^0=0$ is proportional to $h(y)$. 
For small values of $m_H$ we can correct this to $\mathcal O(m_H^2)$  which yields the corresponding (properly normalized) wave function
\be
\xi_H(y)=\sqrt{\frac{ky_1}{Z}}\frac{h(y)}{h(y_1)}e^{A(y_1)}\left[1-m_H^2\left(\int_0^y e^{2A}\frac{\Omega}{\Omega'}
+\int_0^{y_1} e^{2A}\frac{\Omega}{\Omega'}(\Omega-1)\right)\right]\,
\ ,
\label{xiH}
\ee
where the function $\Omega(y)$ was defined in Eq.~(\ref{omegafunction}). The true value of $m_H$ is of course determined by the boundary conditions.
With the usual choice of the boundary potentials
\be\lambda^0=M_0 |H|^2
\,,\qquad
-\lambda^1=-M_1 |H|^2+\gamma |H|^4
\ ,
\label{boundpot}
\ee
and after using the BC (\ref{HiggsBC}) and (\ref{bckBC}), as well as the definition of $Z$~\footnote{As already mentioned above, the quantity $\sqrt{Z}$ can be viewed as a wave function renormalization in the effective theory, obtained by integrating over the true zero mode $\xi(x,y)=\mathcal H(x)h(y)/h_1$ to obtain
$-\mathcal L_{\rm eff}=e^{-2A_1}k^{-1}Z|D_\mu \mathcal  H|^2+e^{-4A_1}[(h'_1/h_1-M_1)|\mathcal H|^2+\gamma|\mathcal H|^4]$.
}, Eq.~(\ref{Xgeneric}), one obtains for the light Higgs mass
\be
m_H^2
=(kZ)^{-1}2\left(M_1-\frac{h'_1}{h_1}\right)\rho^2
\ .
\label{mH}
\ee

Using the definition of the $WW\xi_n$ coupling~\cite{CGQ3}
\be
h_{WW\xi_n}=\frac{g}{y_1}\int_0^{y_1} dy\,e^{-A(y)}M_A(y)f_0^2(y)\xi_n(y)
\ee
and the wave function (\ref{xiH}), one can deduce that
\be
h_{WW H}=h_{WW H}^{SM}\left[1-\mathcal O(m_H^2/m_{KK}^2,m_W^2/m_{KK}^2)\right]
,
\ee
so the light Higgs unitarizes the theory in a similar way to the SM Higgs.

\ 

{\it 3. \underline{RS model}} 

The previous formalism can be applied to any particular 5D model. The simplest and best known case is the RS model where the space is a slice of AdS space, with metric $A(y)=k y$. Assuming an exponential background for the Higgs field as
\be
h(y)=h(y_1) e^{a k( y-y_1)}
,
\label{hbck}
\ee
the $T$ and $S$ parameters can be readily computed from Eqs.~(\ref{T}) and (\ref{S}) yielding
\bea
\alpha(m_Z) T_{RS}&=& s^2_W \frac{m_Z^2}{\rho^2}(ky_1)
\frac{(a-1)^2}{a(2a-1)}+\dots\label{TRS}
,
\\
\alpha(m_Z)S_{RS}&=&2 s_W^2c_W^2
\frac{m_Z^2}{\rho^2}\frac{a^2-1}{a^2}+\dots
,
\label{SRS}
\eea
where the ellipses indicate subleading corrections in the large volume $ky_1$ and $\rho=k\exp(-ky_1)$. 
In the holographic dual, the quantity $a$ corresponds to the dimension of the Higgs condensate and we demand $a\geq 2$ in order to solve the hierarchy problem.
In the RS model the function $F$ relating $m_{KK}$ to $\rho$ is given by $F(A_{RS})\simeq 2.4$. These expressions agree precisely with the recent result in Ref.~\cite{Round:2010kj}. One can see from (\ref{TRS}) that the contribution to the $T$~parameter is volume enhanced while that to the $S$ parameter is not. This translates into a very strong bound on $\rho$ when we compare these expressions with the experimental data (\ref{fit}). In particular for a Higgs localized on the IR brane (which corresponds to the $a\to\infty$ limit) the expression (\ref{TRS}) and the experimental fit (\ref{fit}) leads to the bound $\rho>4.3$ TeV (and to the bound on the gauge boson KK-masses $m_{KK}>10.4$ TeV) at the 95\% CL. A quick glance at Eq.~(\ref{TRS}) shows that for a bulk Higgs the previous bounds are alleviated. In particular for the case $a=2$ they are lowered by a factor $\sqrt{3}$ which leads to the bounds $\rho>2.5$ TeV or $m_{KK}>6.0$ TeV. In order to decrease the actual bounds on KK-masses, it has been proposed in the literature~\cite{Agashe:2003zs} to gauge an extra 
$SU(2)_R\times U(1)_{B-L}$
symmetry in the bulk  such that there is a residual custodial symmetry which protects the $T$~parameter. In that case the experimental bound on the $S$ parameter (\ref{fit}) translates into the bound $\rho>1.4$ TeV ($m_{KK}>3.3$ TeV) for the case of a localized Higgs and  $\rho>1.2$ TeV ($m_{KK}>2.8$ TeV) for the case of a delocalized Higgs with $a=2$.

%\ 

{\it 4. \underline{The model}} 

 In the rest of this Letter we will explore an alternative solution to the problem of the $T$ parameter based on singular metrics. We will see that the combined effect of the Higgs delocalization and the strength and vicinity of the IR brane to the singularity makes the $T$ parameter comparable to the $S$~parameter.
We will then consider the metric singular at $y=y_s$ \cite{Cabrer:2009we}
\be
A(y)=ky-\frac{1}{\nu^2}\log\left(1-\frac{y}{y_s}  \right)\,,
\label{metrica}
\ee
where $\nu>0$ is a real parameter. In the $\nu\to\infty$ limit it coincides with the RS metric. Moreover, it is AdS near the UV brane and has a singularity at $y_s=y_1+\Delta$, outside the slice between the UV and IR branes and at a distance $\Delta$ from the IR brane. We will also assume that the Higgs background is exponential and given by Eq.~(\ref{hbck}). The gravitational setup which can provide such background will be considered later on. For the moment we will just analyze the impact of the departure from AdS in the IR region in the EWPT parameters $T$ and $S$. This departure is controlled by the parameters $\nu$ and $\Delta$. The smaller $\nu$ and/or $\Delta$ the more the IR brane feels the nearby singularity.

\begin{figure}[tb]
\centering
\begin{psfrags}
%%% PSfrag %%%
\def\PFGstripminus-#1{#1}%
\def\PFGshift(#1,#2)#3{\raisebox{#2}[\height][\depth]{\hbox{%
  \ifdim#1<0pt\kern#1 #3\kern\PFGstripminus#1\else\kern#1 #3\kern-#1\fi}}}%
\providecommand{\PFGstyle}{\small}%
\psfrag{a2}[cc][cc]{\PFGstyle \footnotesize $\text{a=2}$}%
\psfrag{a3}[cc][cc]{\PFGstyle \footnotesize $\text{a=3}$}%
\psfrag{aInfinity}[cc][cc]{\PFGstyle \footnotesize $\text{a=$\infty $}$}%
\psfrag{Nu}[tc][bc]{\PFGstyle $\nu $}%
\psfrag{S08}[tc][tc]{\PFGstyle $0.8$}%
\psfrag{S10}[tc][tc]{\PFGstyle $1.0$}%
\psfrag{S12}[tc][tc]{\PFGstyle $1.2$}%
\psfrag{S14}[tc][tc]{\PFGstyle $1.4$}%
\psfrag{S16}[tc][tc]{\PFGstyle $1.6$}%
\psfrag{S18}[tc][tc]{\PFGstyle $1.8$}%
\psfrag{S20}[tc][tc]{\PFGstyle $2.0$}%
\psfrag{TTAdS}[bc][tc]{\PFGstyle $T/T_{\mathrm{RS}}$}%
\psfrag{W02}[cr][cr]{\PFGstyle $0.2$}%
\psfrag{W04}[cr][cr]{\PFGstyle $0.4$}%
\psfrag{W06}[cr][cr]{\PFGstyle $0.6$}%
\psfrag{W08}[cr][cr]{\PFGstyle $0.8$}%
\psfrag{W10}[cr][cr]{\PFGstyle $1.0$}%
\psfrag{W12}[cr][cr]{\PFGstyle $1.2$}%
\psfrag{W14}[cr][cr]{\PFGstyle $1.4$}%
\psfrag{W16}[cr][cr]{\PFGstyle $1.6$}%
\psfrag{W18}[cr][cr]{\PFGstyle $1.8$}%
%%% PSfrag %%%
\includegraphics[width=0.48\textwidth]{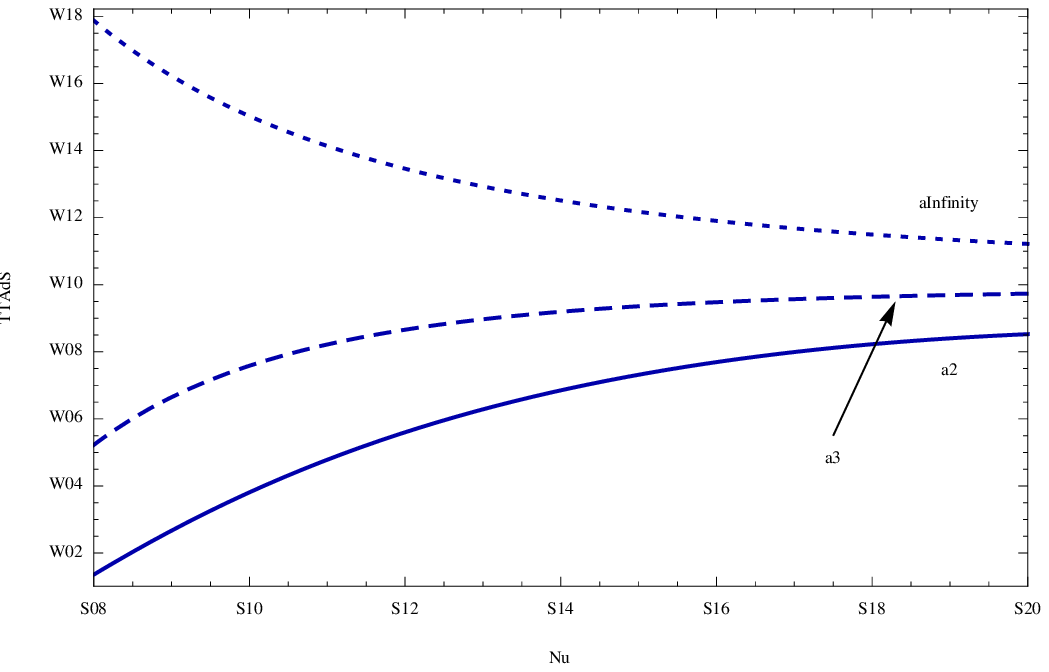}
\end{psfrags}
~~
\begin{psfrags}
%%% PSfrag %%%
\def\PFGstripminus-#1{#1}%
\def\PFGshift(#1,#2)#3{\raisebox{#2}[\height][\depth]{\hbox{%
  \ifdim#1<0pt\kern#1 #3\kern\PFGstripminus#1\else\kern#1 #3\kern-#1\fi}}}%
\providecommand{\PFGstyle}{\small}%
\psfrag{a2}[cc][cc]{\PFGstyle \footnotesize$\text{a=2}$}%
\psfrag{a3}[cc][cc]{\PFGstyle \footnotesize $\text{a=3}$}%
\psfrag{aInfinity}[cc][cc]{\PFGstyle \footnotesize $\text{a=$\infty $}$}%
\psfrag{Nu}[tc][bc]{\PFGstyle $\nu $}%
\psfrag{S08}[tc][tc]{\PFGstyle $0.8$}%
\psfrag{S10}[tc][tc]{\PFGstyle $1.0$}%
\psfrag{S12}[tc][tc]{\PFGstyle $1.2$}%
\psfrag{S14}[tc][tc]{\PFGstyle $1.4$}%
\psfrag{S16}[tc][tc]{\PFGstyle $1.6$}%
\psfrag{S18}[tc][tc]{\PFGstyle $1.8$}%
\psfrag{S20}[tc][tc]{\PFGstyle $2.0$}%
\psfrag{SSAdS}[bc][tc]{\PFGstyle $S/S_{\mathrm{RS}}$}%
\psfrag{W03}[cr][cr]{\PFGstyle $0.3$}%
\psfrag{W04}[cr][cr]{\PFGstyle $0.4$}%
\psfrag{W05}[cr][cr]{\PFGstyle $0.5$}%
\psfrag{W06}[cr][cr]{\PFGstyle $0.6$}%
\psfrag{W07}[cr][cr]{\PFGstyle $0.7$}%
\psfrag{W08}[cr][cr]{\PFGstyle $0.8$}%
\psfrag{W09}[cr][cr]{\PFGstyle $0.9$}%
\psfrag{W10}[cr][cr]{\PFGstyle $1.0$}%
%%% PSfrag %%%
\includegraphics[width=0.48\textwidth]{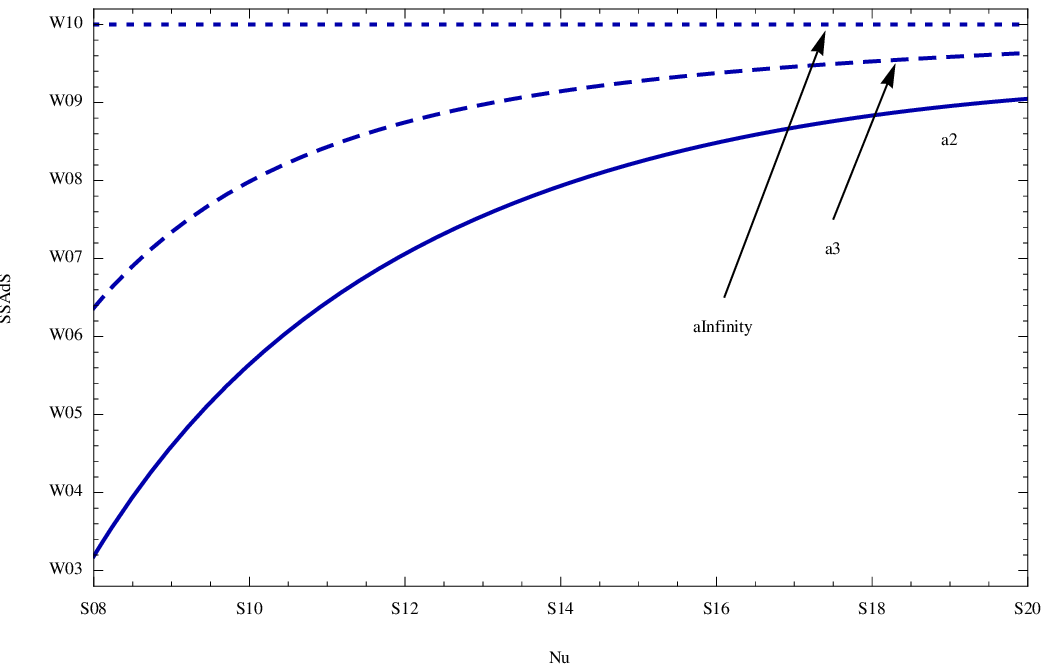}
\end{psfrags}
\caption{\it Plots of the T (left panel) and S (right panel) parameters as a function of $\nu$ for different values of $a$, $k\Delta = 1$ and keeping the masses of the first heavy KK-mode of gauge bosons constant. The parameters are normalized to $1$ when $\nu=\infty$ (i.e. the RS case). }
\label{fig1}
\end{figure}
As we can see in Fig.~\ref{fig1}, the singularity affects both the $T$ and the $S$ parameter~\footnote{In this letter we will consider the independent bounds on $S$ and $T$. Notice that for $T/S\sim +1$, by reducing $m_{KK}$ from infinity to finite values, we move in the $T-S$ plane from the origin along the major (long) axis of the ellipse, taking advantage of that particular correlation in the fit.}. In the left [right] panel we plot the ratio $T/T_{RS}$ [$S/S_{RS}$] where $T$ [$S$] is the parameter obtained from Eq.~(\ref{T}) [Eq.~(\ref{S})], as a function of $\nu$ and for different values of $a$, and $T_{RS}$ [$S_{RS}$] is the corresponding parameter in the RS-model as given by Eq.~(\ref{TRS}) [Eq.~(\ref{SRS})]. In the plots we consider the same value of $m_{KK}$ in both theories. For the metric (\ref{metrica}) the relation between $m_{KK}$ and $\rho$ is given by the function $F(\nu)$ which is a monotonically decreasing function and reproduces the RS-result in the limit $\nu\to\infty$. An approximate fit for the function $F(\nu)$ and values $\nu\gtrsim 0.8$ with $k\Delta=1$ is provided by
\be
F(\nu)\simeq 2.44+\frac{1.71}{\nu^2}\ .
\ee
Since the observables are quadratically dependent on $m_{KK}$ this means that one can read the corresponding reduction on the value of $m_{KK}$, with respect to the RS-bound, by taking the square root of the vertical axis in the respective plot of Fig.~\ref{fig1}. 

We can see from the left panel of Fig.~\ref{fig1} that there is no reduction on the $T$ parameter for a localized Higgs in agreement with the general results of Refs.~\cite{Delgado:2007ne,Archer:2010hh}. On the other hand for a bulk Higgs the corresponding reduction in the $T$ and $S$ parameters can be qualitatively understood in the limit of large wave-function renormalization $Z$ from the approximate expressions in Eqs.~(\ref{Tapp}) and (\ref{Sapp}) in the limit of large volume suppression. To see this better let us integrate out the KK modes of the gauge bosons.
This produces dimension-six operators quadratic in the Higgs and fermion electroweak currents. 
The operator quadratic in the Higgs currents contributes to $T$, while the operator proportional to the product of one Higgs and one fermionic current contributes to $S$ after going to the "oblique basis" \cite{Davoudiasl:2009cd}. Notice that the coupling of the KK modes to the Higgs current is proportional to $Z^{-1}$. The question is whether there are other effects that influence these couplings.
They can be calculated as integrals over KK wave functions, which can be obtained only numerically. 
However, we can easily evaluate the quantity $I$ appearing in our expressions for $S$ and $T$. 
% in terms of the gauge boson wave functions at the UV brane as \cite{CGQ3}
%
%\be
%\frac{I}{ky_1\rho^2}=\sum_{n\geq 1}\left(\frac{f_A^n(0)}{m_{f_A}^n}\right)^2.
%\label{GFKK}
%\ee
%
Numerically it turns out that $I\, m_{KK}^2/\rho^2$ is fairly constant (in fact, it shows some additional mild suppression) over the interesting range of $\nu$ and $\Delta$, and hence it is reasonable to expect that the couplings are mainly affected by the parameters $\nu$, $\Delta$ and $a$ through the dependence on $Z$.
It is noteworthy that the IR dimension of the Higgs condensate in the holographic 4D dual is reduced with respect to its UV conformal value [$\dim (\mathcal O_H^{UV})=a$] leading to the enhanced Z factors responsible for the suppression of the $S$ and $T$ parameters. Indeed, identifying the logarithm of the RG scale with the metric as $\log Q=-A(y)$, the IR dimension of the Higgs can be expressed as 
\be
\dim(\mathcal O_H^{IR})=\frac{a}{1+\frac{1}{k\Delta\nu^2}}.
\ee
Moreover, the number of colors $N_c$ of the effective theory is also reduced in the IR region since the curvature increases along the extra dimension due to the presence of the singularity~\footnote{This phenomenon has been extensively studied in warped throat geometries with a singularity which appear e.g.~in the Klebanov-Tseytlin solution of type IIB string constructions~\cite{Klebanov:2000nc,Brummer:2005sh}.}:
\be
\frac{N_c^{IR}}{N_c^{UV}}=\left(\frac{R^{IR}}{R^{UV}}\right)^{-3/4}=
\left[\left(1+\frac{1}{k\Delta\nu^2}\right)^2-\frac{2}{5}\frac{1}{(k\Delta\nu)^2}\right]^{-3/4}.
\ee
It is interesting that both effects, small $N_c$ and $\dim (\mathcal O_H^{IR})<2$, also play a major role in reducing $S$ and $T$ in the conformal technicolor model proposed in Ref.~\cite{Luty:2004ye}, the main difference being that our proposal implies a strong deformation of the conformal theory in the IR. 
More details will be presented elsewhere~\cite{CGQ3}.

A 5D setup leading to the background (\ref{metrica}) and (\ref{hbck}) can be easily obtained by using the formalism of Ref.~\cite{DeWolfe:1999cp},
where first-order gravitational EOM and the bulk potential can be obtained from a superpotential. We will introduce on top of the SM Higgs field $H$ a scalar field $\phi$ which will generate a singularity at $y=y_s$, with the superpotential $W(\phi,H)$ related to the scalar potential by
\be
 V(\phi,h) \equiv \frac{1}{2} 
 			\left[ 
				\left( \frac{\partial W}{\partial \phi} \right)^2 
				+
				\left( \frac{\partial W}{\partial h} \right)^2 
			\right]
			- 
			\frac{1}{3} W(\phi,h)^2 
	\ .
	\label{VW}
\ee
Using this ansatz the bulk EOMs can be written as simple first-order differential equations 
\be
A'(y) = \frac{1}{6} W(\phi(y),h(y)) ,\quad
\phi'(y) = \partial_\phi W(\phi,h),\quad 
h'(y) = \partial_h W(\phi,h)\,.
 \label{metricEOM}
\ee
In terms of the boundary potentials $\lambda^\alpha(\phi,h)
$, the boundary conditions are
\be
A'(y_\alpha)=\left.\frac{2}{3}\lambda^\alpha(\phi,h)\right|_{y=y_\alpha},\quad
\phi'(y_\alpha)=\left.\frac{\partial\lambda^\alpha}{\partial\phi}\right|_{y=y_\alpha},\quad
h'(y_\alpha)=\left.\frac{\partial\lambda^\alpha}{\partial h}\right|_{y=y_\alpha}\ .
\label{BCs}
\ee
 
We postulate the superpotential $W(\phi,H)=W_\phi(\phi)+W_H(h)$ where
\be
W_\phi(\phi)=6k(1+b e^{\nu\phi/\sqrt{6}}),\quad W_H(h)=\frac12 a k h^2 ,
\label{superp}
\ee
and $a$ and $b$ are real arbitrary parameters. This leads to the background configuration (\ref{hbck}) and~\cite{Gubser:2000nd}
\bea
\phi(y)&=&-\frac{\sqrt{6}}{\nu}\log[\nu^2 b k(y_s-y)]\label{phi}
\ ,
\\
A(y)&=&ky-\frac{1}{\nu^2}\log\left(1-\frac{y}{y_s}\right)+\frac{1}{24}(h^2(y)-h^2(0))
\ ,
\label{A}
\end{eqnarray}
where we are using the normalization $A(0) =  0$.  Notice that the Higgs contributions to the metric near the UV brane and near the
singularity at $y=y_s$ are overwhelmed by that of the $\phi$
background. Thus, the metric (\ref{A}) agrees to all practical purposes with (\ref{metrica}). 

Using the superpotential formalism amounts to some fine-tuning among the different coefficients of the bulk potential, unless they are protected by some underlying 5D supergravity~\cite{DeWolfe:1999cp}. The quadratic Higgs term which is generated by (\ref{VW}) can be written as
$k^2\left[a(a-4)-4 a b e^{\nu\phi/\sqrt{6}} \right]|H|^2$
and the coefficients of the two operators $|H|^2$ and $e^{\nu\phi/\sqrt{6}}|H|^2$ can be considered as independent parameters~\footnote{Of course the coefficients of the operators not involving  the Higgs field remain fine-tuned.}. However, since the parameter $b$ can be traded by a global shift in the value of the $\phi$ field, or in particular by a shift in its value at the UV brane $\phi_0$, for simplicity we will fix its value to $b=1$ hereafter.

We assume that the brane dynamics $\lambda^\alpha_\phi$ fixes the values of the field $\phi$ at \mbox{$\phi=\phi_0,\,\phi_1$} on the UV and IR branes respectively. The inter-brane distance $y_1$, as well as the location of the singularity at $y_s$ and the warp factor $A(y_1)$, are related to the values of the field $\phi_\alpha$ at the branes by the following expressions:
\begin{eqnarray}
ky_1&=&\frac{1}{\nu^2}\left[e^{-\nu \phi_0/\sqrt{6}}-e^{-\nu \phi_1/\sqrt{6}}  \right], \quad k\Delta=\frac{1}{\nu^2}e^{-\nu \phi_1/\sqrt{6}}
\ ,
\nonumber\\
A(y_1)&\simeq&k y_1+\frac{1}{\nu}(\phi_1-\phi_0)/\sqrt{6}
\ ,
\end{eqnarray}
which shows that the required large hierarchy can be naturally 
fixed with values of the fields $\phi_1\gtrsim \phi_0$, $\phi_0<0$ and of order one.
Moreover, the strict soft-wall configuration~\cite{Cabrer:2009we} corresponds to the limit $\phi_1\gg 1,\, y_1\to y_s$. 
As for the brane potentials $\lambda^\alpha_H$, they are given by Eq.~(\ref{boundpot}). The BC (\ref{BCs}) at the UV brane imply~\footnote{This apparent fine tuning is an artifact of the first order formulation and has no observable physical consequences for $a>2$. Indeed for $a>2$ the parameter $M_0$ has no physical impact neither on the background nor in the spectrum.} $M_0-ak=0$, while the BC at $y_1$ fixes the value of the Higgs background at the IR brane $h_1$ as $\gamma h_1^2=M_1-ak$ and thus triggers electroweak symmetry breaking. Also note that due to its exponential dependence on $\phi_1$, $\Delta$ can be small or, in other words, the IR brane naturally occurs very close to the singularity.

\begin{figure}[tb]
\centering
\begin{psfrags}
%%% PSfrag %%%
\def\PFGstripminus-#1{#1}%
\def\PFGshift(#1,#2)#3{\raisebox{#2}[\height][\depth]{\hbox{%
  \ifdim#1<0pt\kern#1 #3\kern\PFGstripminus#1\else\kern#1 #3\kern-#1\fi}}}%
\providecommand{\PFGstyle}{\small}%
\psfrag{BoxmfTeV}[bc][bc]{\PFGstyle $m_{\rm KK}~(\mathrm{TeV})$}%
\psfrag{Nu}[tc][bc]{\PFGstyle $\nu $}%
\psfrag{S08}[tc][tc]{\PFGstyle $0.8$}%
\psfrag{S10}[tc][tc]{\PFGstyle $1.0$}%
\psfrag{S12}[tc][tc]{\PFGstyle $1.2$}%
\psfrag{S14}[tc][tc]{\PFGstyle $1.4$}%
\psfrag{S16}[tc][tc]{\PFGstyle $1.6$}%
\psfrag{S18}[tc][tc]{\PFGstyle $1.8$}%
\psfrag{S20}[tc][tc]{\PFGstyle $2.0$}%
\psfrag{S}[cc][cc]{\PFGstyle $S$}%
\psfrag{T}[cc][cc]{\PFGstyle $T$}%
\psfrag{W1}[cr][cr]{\PFGstyle $1$}%
\psfrag{W2}[cr][cr]{\PFGstyle $2$}%
\psfrag{W3}[cr][cr]{\PFGstyle $3$}%
\psfrag{W4}[cr][cr]{\PFGstyle $4$}%
\psfrag{W5}[cr][cr]{\PFGstyle $5$}%
\psfrag{W6}[cr][cr]{\PFGstyle $6$}%
%%% PSfrag %%%
\includegraphics[width=0.49\textwidth]{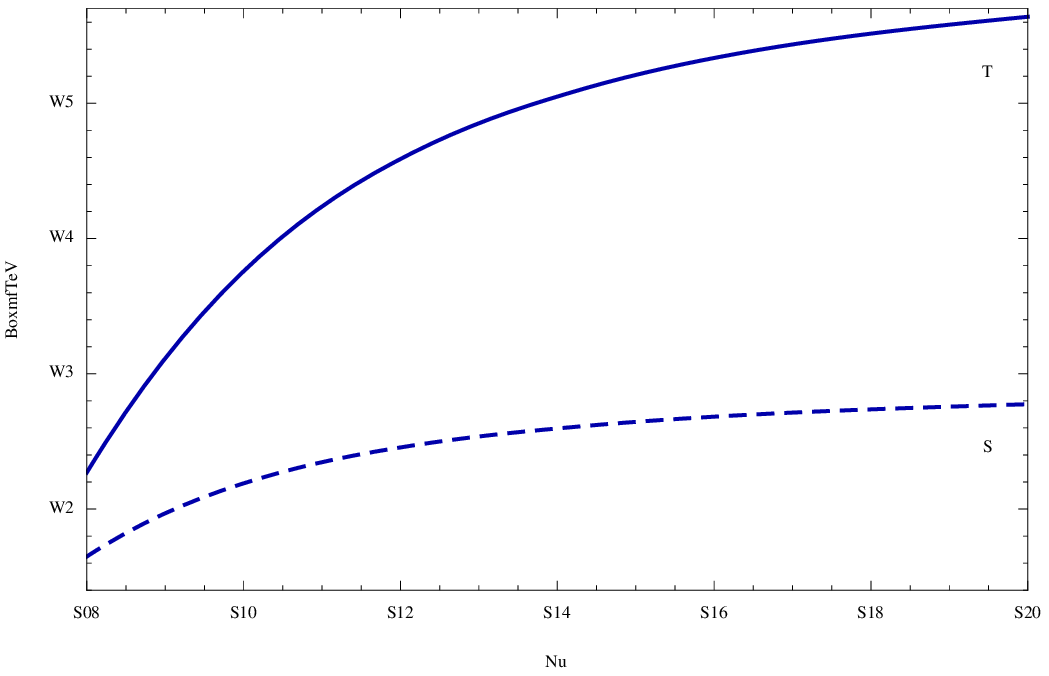}
\end{psfrags}
\begin{psfrags}
%%% PSfrag %%%
\def\PFGstripminus-#1{#1}%
\def\PFGshift(#1,#2)#3{\raisebox{#2}[\height][\depth]{\hbox{%
  \ifdim#1<0pt\kern#1 #3\kern\PFGstripminus#1\else\kern#1 #3\kern-#1\fi}}}%
\providecommand{\PFGstyle}{\small}%
\psfrag{BoxmfTeV}[bc][bc]{\PFGstyle }%
\psfrag{kCDelta}[tc][tc]{\PFGstyle $\text{k$\Delta $}$}%
\psfrag{S02}[tc][tc]{\PFGstyle $0.2$}%
\psfrag{S04}[tc][tc]{\PFGstyle $0.4$}%
\psfrag{S06}[tc][tc]{\PFGstyle $0.6$}%
\psfrag{S08}[tc][tc]{\PFGstyle $0.8$}%
\psfrag{S10}[tc][tc]{\PFGstyle $1.0$}%
\psfrag{S}[cc][cc]{\PFGstyle $S$}%
\psfrag{T}[cc][cc]{\PFGstyle $T$}%
\psfrag{W1}[cr][cr]{\PFGstyle $1$}%
\psfrag{W2}[cr][cr]{\PFGstyle $2$}%
\psfrag{W3}[cr][cr]{\PFGstyle $3$}%
\psfrag{W4}[cr][cr]{\PFGstyle $4$}%
\psfrag{W5}[cr][cr]{\PFGstyle $5$}%
\psfrag{W6}[cr][cr]{\PFGstyle $6$}%
%%% PSfrag %%%
\includegraphics[width=0.49\textwidth]{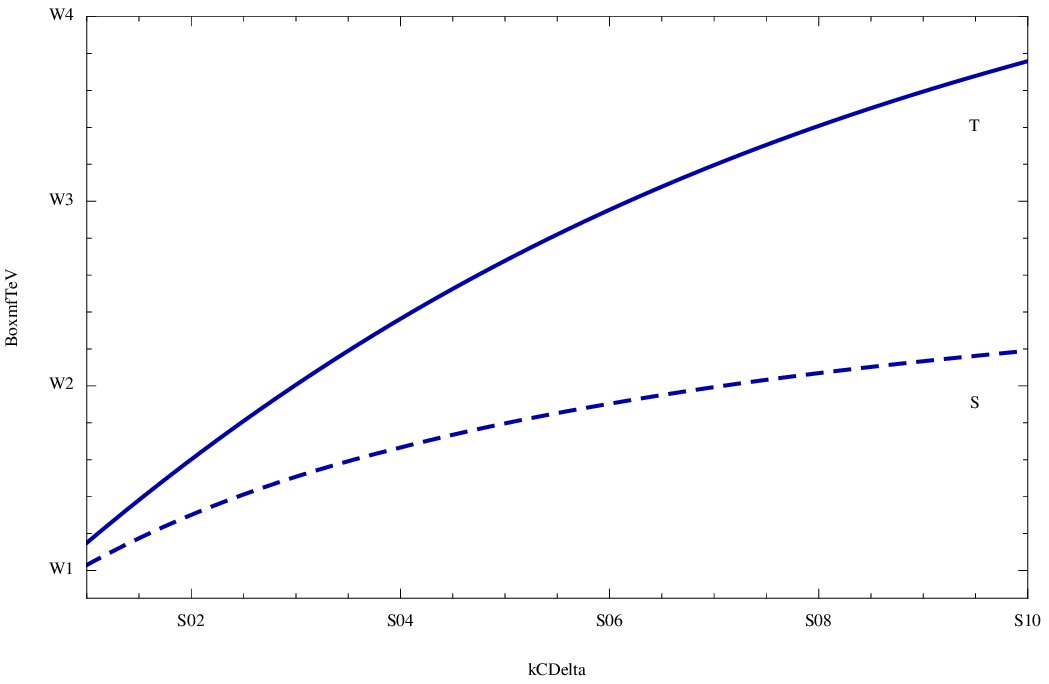}
\end{psfrags}
\caption{\it Plots of the 95\% CL lower bounds on the first KK-mode mass of electroweak gauge bosons from experimental bounds on the $T$ and $S$ parameters for $A(y_1)=35$ and $a=2$ as a function of $\nu$ (left panel, with $k\Delta=1$) and as a function of $k\Delta$ (right panel, with $\nu=1$).}
\label{fig2}
\end{figure}
We will now explore numerically the predictions of the model defined by the background (\ref{hbck}), (\ref{phi}) and (\ref{A}). In particular, the KK-modes of the gauge bosons satisfy Eq.~(\ref{eq.f}). This equation can be solved numerically and the free parameters are $(y_1,\Delta,a,\nu)$. 
%%%%
We will fix the Planck/weak hierarchy by imposing $A(y_1)\simeq 35$. This establishes a functional relation $y_1=y_1(\Delta,\nu)$ by which $y_1$ increases~\footnote{The fact that $y_1$ is an increasing function of $\Delta$ and $\nu$ helps lowering the bounds on $\rho$ from the $T$ parameter for small $\nu$ and $\Delta$, since the $T$ parameter is volume enhanced [see Eq.~\eqref{Tapp}]. However this effect is small in the parameter region shown in the figures.} with $\Delta$ and $\nu$. The rest of parameters are free and they have a clear physical meaning. The parameter $a$ indicates the departure from a localized Higgs, which is the limit $a\to\infty$. The parameter $\nu$ indicates the departure from the RS case, which is the limit $\nu\to\infty$. The parameter $\Delta$ indicates the distance between the IR brane and the singularity. All the effects that we consider are enhanced when $a$, $\nu$ and $\Delta$ decrease. 
%%%%%
%
We plot in Fig.~\ref{fig2} the lower bounds obtained on the mass $m_{KK}$ of the first KK-mode for the electroweak gauge bosons $W,Z$ and $\gamma$ from the 95\% CL bounds on the $T$ and $S$ parameters of Eq.~(\ref{fit}). On the left plot we see the dependence on the $\nu$ parameter for $a=2$ and $k\Delta=1$. We see that for $\nu\gg 1$ the bounds go to the RS bounds for a bulk Higgs which are around $m_{KK}>$ 6.0 TeV and 2.8 TeV from the $T$ and $S$ parameters respectively. On the right plot we see the influence on the bounds of the vicinity to the singularity. In summary, for $\nu\lesssim 1$ and/or $k\Delta\lesssim 1$ there are regions where the bounds are $m_{KK}=\mathcal O( 1-3)$~TeV. %These regions are small, but they are not fine-tuned.

\begin{figure}[tb]
\centering
\begin{psfrags}
%%% PSfrag %%%
\def\PFGstripminus-#1{#1}%
\def\PFGshift(#1,#2)#3{\raisebox{#2}[\height][\depth]{\hbox{%
  \ifdim#1<0pt\kern#1 #3\kern\PFGstripminus#1\else\kern#1 #3\kern-#1\fi}}}%
\providecommand{\PFGstyle}{\small}%
\psfrag{BoxmfTeV}[bc][bc]{\PFGstyle $m_{\rm KK}~(\mathrm{TeV})$}%
\psfrag{gauge}[cc][cc]{\PFGstyle \footnotesize $\text{gauge}$}%
\psfrag{Higgs}[cc][cc]{\PFGstyle \footnotesize $\text{Higgs}$}%
\psfrag{Nu}[tc][bc]{\PFGstyle $\nu $}%
\psfrag{S08}[tc][tc]{\PFGstyle $0.8$}%
\psfrag{S10}[tc][tc]{\PFGstyle $1.0$}%
\psfrag{S12}[tc][tc]{\PFGstyle $1.2$}%
\psfrag{S14}[tc][tc]{\PFGstyle $1.4$}%
\psfrag{S16}[tc][tc]{\PFGstyle $1.6$}%
\psfrag{S18}[tc][tc]{\PFGstyle $1.8$}%
\psfrag{S20}[tc][tc]{\PFGstyle $2.0$}%
\psfrag{W1}[cr][cr]{\PFGstyle $1$}%
\psfrag{W2}[cr][cr]{\PFGstyle $2$}%
\psfrag{W3}[cr][cr]{\PFGstyle $3$}%
\psfrag{W4}[cr][cr]{\PFGstyle $4$}%
\psfrag{W5}[cr][cr]{\PFGstyle $5$}%
\psfrag{W6}[cr][cr]{\PFGstyle $6$}%
\psfrag{W7}[cr][cr]{\PFGstyle $7$}%
\psfrag{W8}[cr][cr]{\PFGstyle $8$}%
%%% PSfrag %%%
\includegraphics[width=0.49\textwidth]{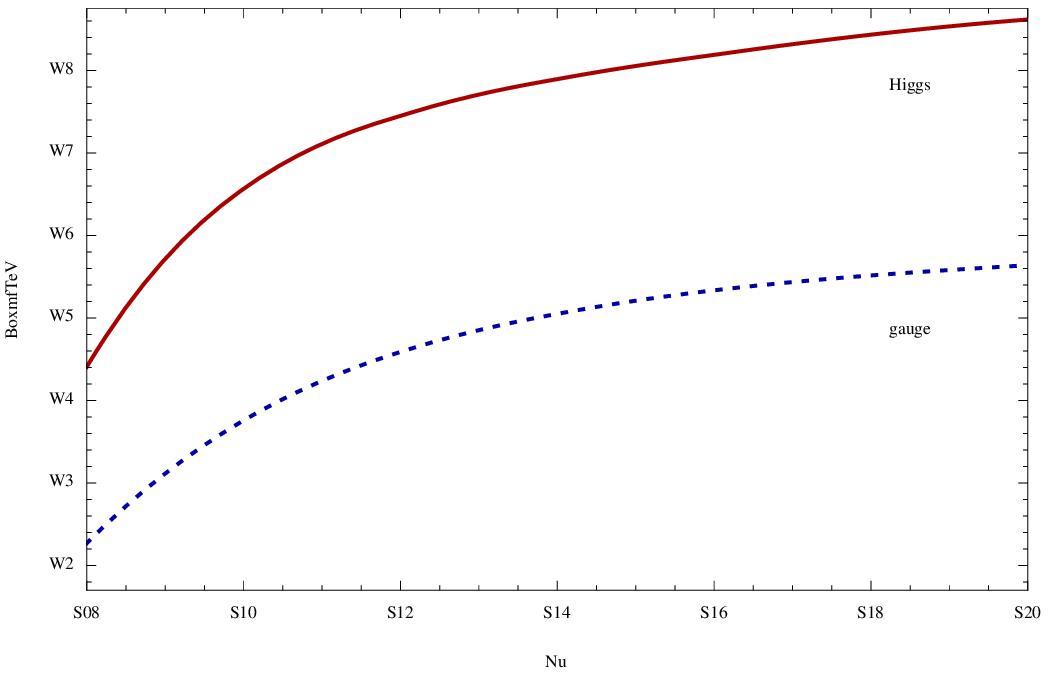}
\end{psfrags}
\begin{psfrags}
%%% PSfrag %%%
\def\PFGstripminus-#1{#1}%
\def\PFGshift(#1,#2)#3{\raisebox{#2}[\height][\depth]{\hbox{%
  \ifdim#1<0pt\kern#1 #3\kern\PFGstripminus#1\else\kern#1 #3\kern-#1\fi}}}%
\providecommand{\PFGstyle}{\small}%
\psfrag{BoxmfTeV}[bc][bc]{\PFGstyle }%
\psfrag{gauge}[cc][cc]{\PFGstyle \footnotesize $\text{gauge}$}%
\psfrag{Higgs}[cc][cc]{\PFGstyle \footnotesize $\text{Higgs}$}%
\psfrag{kCDelta}[tc][tc]{\PFGstyle $\text{k$\Delta $}$}%
\psfrag{S02}[tc][tc]{\PFGstyle $0.2$}%
\psfrag{S04}[tc][tc]{\PFGstyle $0.4$}%
\psfrag{S06}[tc][tc]{\PFGstyle $0.6$}%
\psfrag{S08}[tc][tc]{\PFGstyle $0.8$}%
\psfrag{S10}[tc][tc]{\PFGstyle $1.0$}%
\psfrag{W0}[cr][cr]{\PFGstyle $0$}%
\psfrag{W1}[cr][cr]{\PFGstyle $1$}%
\psfrag{W2}[cr][cr]{\PFGstyle $2$}%
\psfrag{W3}[cr][cr]{\PFGstyle $3$}%
\psfrag{W4}[cr][cr]{\PFGstyle $4$}%
\psfrag{W5}[cr][cr]{\PFGstyle $5$}%
\psfrag{W6}[cr][cr]{\PFGstyle $6$}%
\psfrag{W7}[cr][cr]{\PFGstyle $7$}%
%%% PSfrag %%%
\includegraphics[width=0.49\textwidth]{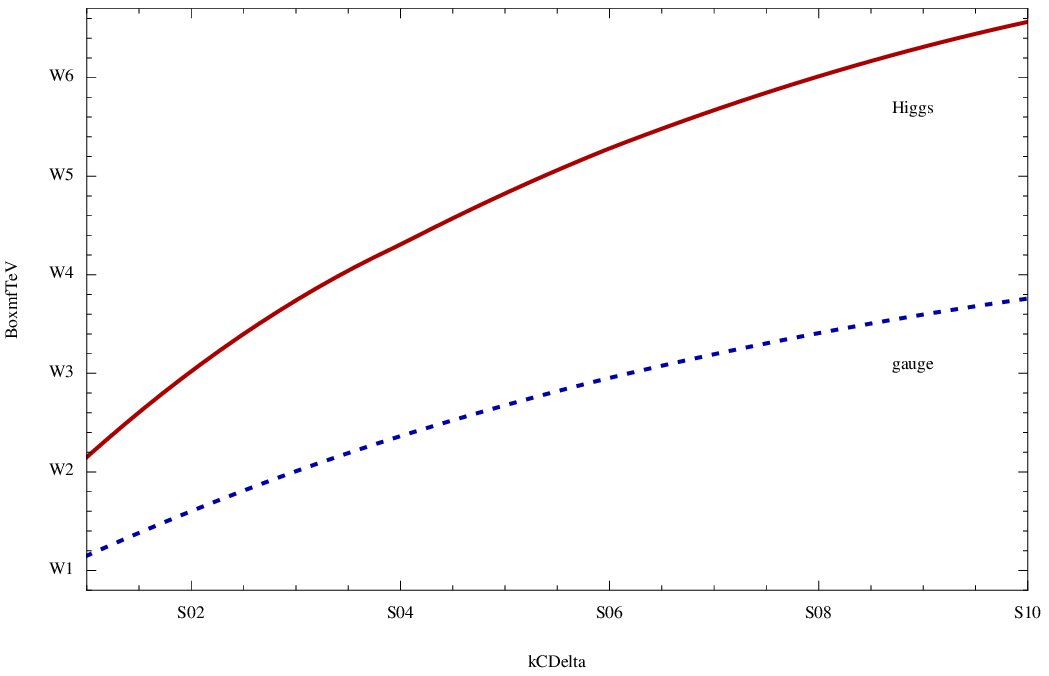}
\end{psfrags}
\caption{\it Plots of the 95\% CL lower bounds on the first KK-mode mass of the Higgs boson from experimental bounds on the $T$ parameter for $A(y_1)=35$ and $a=2$ as a function of $\nu$ (left panel, with $k\Delta=1$) and as a function of $k\Delta$ (right panel, with $\nu=1$). For comparison the corresponding bounds on the KK-gauge boson masses are shown.}
\label{fig3}
\end{figure}
Higgs fluctuations satisfy the EOM (\ref{Higgsbulk}) and the BC (\ref{HiggsBC}). Plots based on numerical solutions are shown in Fig.~\ref{fig3}, where
the 95\% CL bounds on the first KK-mode Higgs masses are shown. From Fig.~\ref{fig3} we see that Higgs KK-modes are heavier than gauge boson ones, disfavoring its experimental detection at LHC. Moreover, as we have seen earlier in this Letter, there is a light Higgs eigenstate with wave function given by (\ref{xiH}) and mass eigenvalue given by (\ref{mH}). After using the background metric $A(y)$ in Eq.~(\ref{A}) and the background Higgs $h(y)$ in Eq.~(\ref{hbck}), 
the Higgs mass eigenvalue is given by
\be
\frac{m_H^2}{\rho^2}=\frac{2\mu}{ Z}
\ ,
\\
\ee
\be
Z=e^{2(a-1)\Delta}\Delta^{1-\eta}[2(a-1)]^{-\eta}\Gamma(\eta,2(a-1)\Delta)
\ ,
\label{ultima}
\ee 
where $ \mu\equiv M_1/k-a$, $\eta=1+\frac{2}{\nu^2}$ and $\Gamma$ stands for the incomplete gamma function. Equation (\ref{ultima}) provides a measure of the required fine-tuning to get a light Higgs. In fact for $\mu=0$ (which amounts to an infinite fine-tuning and no EWSB) the Higgs is massless. In general the smallness of $ \mu$ is a measure of the degree of fine-tuning required to achieve a light Higgs. In fact, the fine-tuning is smaller than in the RS-model because the prefactor of $ \mu$ in (\ref{ultima}) is smaller than one. For instance, for $a=2$, $\nu=0.9$ and $k\Delta=0.8$  the lower bound on $\rho$ from the $T$~parameter is $\rho\simeq 0.49$ TeV and $m_H\simeq \, 470\, \sqrt{\mu}$ GeV, providing a light Higgs for $\mu=\mathcal O(10^{-1})$.

\

{\it5. \underline{Conclusions}} 

In this Letter we have presented an alternative mechanism to the introduction of an extra gauge symmetry to suppress the $T$~parameter in warped extra dimensional models. The mechanism is based on the introduction of a metric that is nearly AdS in the UV region but departs from AdS in the IR region, towards where KK-fluctuations are mainly localized, and that has a singularity outside the slice between the UV and IR branes. The two main parameters which control this effect are then the departure of the metric with respect to the AdS metric and the vicinity of the IR brane to the singularity. Depending on these parameters, the lower bounds on the gauge boson KK-mode masses can be as low as around the TeV scale. This low bound should alleviate the little hierarchy problem that arises in warped electroweak breaking models and facilitate experimental detection of these heavy gauge bosons at LHC. The model requires a scalar field $\phi$ propagating in the bulk, which plays the role of a radion stabilizing field, similar to the Goldberger-Wise scalar in the RS theory. We have considered the back-reactions of the $\phi$ and Higgs fields on the metric by means of the superpotential formalism, which requires some fine-tuning in the gravitational part of the bulk potential. The model also contains a naturally light Higgs that unitarizes the 4D theory in much a similar way as the SM Higgs.

\

{\it\underline{Acknowledgments}}\nopagebreak 

Work supported in part by the Spanish Consolider-Ingenio 2010
Programme CPAN (CSD2007-00042) and by CICYT, Spain, under contract
FPA 2008-01430. The work of JAC is supported by the Spanish Ministry of Education through a FPU grant. The work of GG is supported by the ERC Advanced Grant 226371, the ITN
programme PITN- GA-2009-237920 and the IFCPAR CEFIPRA programme 4104-2.
JAC and MQ wish to thank CPTH (\'Ecole Polytechnique, Paris) where part of this work has been done for hospitality. We wish to thank J.~Santiago and E.~Pont\'on for discussions and correspondence on the $S$, $T$ and $U$ parameters in warped models. GG would like to thank Kaustubh Agashe for helpful discussions.


\begin{thebibliography}{99}

\bibitem{RS1}   L.~Randall and R.~Sundrum,
  %``A large mass hierarchy from a small extra dimension,''
  Phys.\ Rev.\ Lett.\  {\bf 83}, 3370 (1999)
  [arXiv:hep-ph/9905221].
  %%CITATION = PRLTA,83,3370;%%

\bibitem{Huber:2000fh}
  S.~J.~Huber and Q.~Shafi,
  %``Higgs mechanism and bulk gauge boson masses in the Randall-Sundrum
  %model,''
  Phys.\ Rev.\  D {\bf 63}, 045010 (2001)
  [arXiv:hep-ph/0005286].
  %%CITATION = PHRVA,D63,045010;%%

\bibitem{Davoudiasl:2009cd}
  H.~Davoudiasl, S.~Gopalakrishna, E.~Ponton and J.~Santiago,
  %``Warped 5-Dimensional Models: Phenomenological Status and Experimental
  %Prospects,''
  New J.\ Phys.\  {\bf 12} (2010) 075011
  [arXiv:0908.1968 [hep-ph]], and references therein.
  %%CITATION = NJOPF,12,075011;%%

\bibitem{Agashe:2003zs}
  K.~Agashe, A.~Delgado, M.~J.~May and R.~Sundrum,
  %``RS1, custodial isospin and precision tests,''
  JHEP {\bf 0308} (2003) 050
  [arXiv:hep-ph/0308036].
  %%CITATION = JHEPA,0308,050;%%

\bibitem{Agashe:2006at}
  K.~Agashe, R.~Contino, L.~Da Rold and A.~Pomarol,
  %``A custodial symmetry for Z b anti-b,''
  Phys.\ Lett.\  B {\bf 641} (2006) 62
  [arXiv:hep-ph/0605341].
  %%CITATION = PHLTA,B641,62;%%

\bibitem{Davoudiasl:2002ua}
  H.~Davoudiasl, J.~L.~Hewett and T.~G.~Rizzo,
  %``Brane localized kinetic terms in the Randall-Sundrum model,''
  Phys.\ Rev.\  D {\bf 68} (2003) 045002
  [arXiv:hep-ph/0212279];
  %%CITATION = PHRVA,D68,045002;%%
  M.~S.~Carena, E.~Ponton, T.~M.~P.~Tait and C.~E.~M.~Wagner,
  %``Opaque branes in warped backgrounds,''
  Phys.\ Rev.\  D {\bf 67} (2003) 096006
  [arXiv:hep-ph/0212307].
  %%CITATION = PHRVA,D67,096006;%%


\bibitem{Delgado:2007ne}
  A.~Delgado and A.~Falkowski,
  %``Electroweak observables in a general 5D background,''
  JHEP {\bf 0705} (2007) 097
  [arXiv:hep-ph/0702234].
  %%CITATION = JHEPA,0705,097;%%

\bibitem{Archer:2010hh}
  P.~R.~Archer and S.~J.~Huber,
  %``Electroweak Constraints on Warped Geometry in Five Dimensions and Beyond,''
  JHEP {\bf 1010} (2010) 032
  [arXiv:1004.1159 [hep-ph]].
  %%CITATION = JHEPA,1010,032;%%

\bibitem{GW}   W.~D.~Goldberger and M.~B.~Wise,
  %``Modulus stabilization with bulk fields,''
  Phys.\ Rev.\ Lett.\  {\bf 83}, 4922 (1999)
  [arXiv:hep-ph/9907447];
  %%CITATION = PRLTA,83,4922;%%
 W.~D.~Goldberger and M.~B.~Wise,
  %``Phenomenology of a stabilized modulus,''
  Phys.\ Lett.\  B {\bf 475}, 275 (2000)
  [arXiv:hep-ph/9911457].
  %%CITATION = PHLTA,B475,275;%%


\bibitem{Gubser:2000nd}
  S.~S.~Gubser,
  %``Curvature singularities: The good, the bad, and the naked,''
  Adv.\ Theor.\ Math.\ Phys.\  {\bf 4}, 679 (2000)
  [arXiv:hep-th/0002160].
  %%CITATION = 00203,4,679;%%
  
\bibitem{AdS/QCD}   A.~Karch, E.~Katz, D.~T.~Son and M.~A.~Stephanov,
  %``Linear Confinement and AdS/QCD,''
  Phys.\ Rev.\  D {\bf 74}, 015005 (2006)
  [arXiv:hep-ph/0602229];
  %%CITATION = PHRVA,D74,015005;%%
B.~Batell and T.~Gherghetta,
  %``Dynamical Soft-Wall AdS/QCD,''
  Phys.\ Rev.\  D {\bf 78}, 026002 (2008) 
  [arXiv:0801.4383 [hep-ph]].
  %%CITATION = PHRVA,D78,026002;%%

 \bibitem{Falkowski1} A.~Falkowski and M.~Perez-Victoria,
  %``Electroweak Breaking on a Soft Wall,''
  JHEP {\bf 0812}, 107 (2008)
  [arXiv:0806.1737 [hep-ph]].
  %%CITATION = JHEPA,0812,107;%%
 \bibitem{Falkowski2} 
  A.~Falkowski and M.~Perez-Victoria,
  %``Holographic Unhiggs,''
  Phys.\ Rev.\  D {\bf 79} (2009) 035005
  [arXiv:0810.4940 [hep-ph]].
  %%CITATION = PHRVA,D79,035005;%%

  \bibitem{Batell2}   B.~Batell, T.~Gherghetta and D.~Sword,
  %``The Soft-Wall Standard Model,''
  Phys.\ Rev.\  D {\bf 78}, 116011 (2008)
  [arXiv:0808.3977 [hep-ph]];
  %%CITATION = PHRVA,D78,116011;%%
%\bibitem{Gherghetta:2009qs}
  T.~Gherghetta and D.~Sword,
  %``Fermion Flavor in Soft-Wall AdS,''
  Phys.\ Rev.\  D {\bf 80} (2009) 065015
  [arXiv:0907.3523 [hep-ph]];
  %%CITATION = PHRVA,D80,065015;%%
%\bibitem{Gherghetta:2010he}
  T.~Gherghetta and N.~Setzer,
  %``On the stability of a soft-wall model,''
  Phys.\ Rev.\  D {\bf 82} (2010) 075009
  [arXiv:1008.1632 [hep-ph]].
  %%CITATION = PHRVA,D82,075009;%%


\bibitem{Delgado:2009xb}
  A.~Delgado and D.~Diego,
  %``Fermion Mass Hierarchy from the Soft Wall,''
  Phys.\ Rev.\  D {\bf 80} (2009) 024030
  [arXiv:0905.1095 [hep-ph]].
  %%CITATION = PHRVA,D80,024030;%%

\bibitem{MertAybat:2009mk}
  S.~Mert Aybat and J.~Santiago,
  %``Bulk Fermions in Warped Models with a Soft Wall,''
  Phys.\ Rev.\  D {\bf 80} (2009) 035005
  [arXiv:0905.3032 [hep-ph]].
  %%CITATION = PHRVA,D80,035005;%%
  
 \bibitem{Cabrer:2009we}
   J.~A.~Cabrer, G.~von Gersdorff and M.~Quiros,
  %``Soft-Wall Stabilization,''
  New J.\ Phys.\  {\bf 12} (2010) 075012
  [arXiv:0907.5361 [hep-ph]].
  %%CITATION = NJOPF,12,075012;%%

\bibitem{Atkins:2010cc}
  M.~Atkins and S.~J.~Huber,
  %``Suppressing Lepton Flavour Violation in a Soft-Wall Extra Dimension,''
  Phys.\ Rev.\  D {\bf 82} (2010) 056007
  [arXiv:1002.5044 [hep-ph]].
  %%CITATION = PHRVA,D82,056007;%%
  
  %\cite{vonGersdorff:2010ht}
\bibitem{vonGersdorff:2010ht}
  G.~von Gersdorff,
  %``From Soft Walls to Infrared Branes,''
  Phys.\ Rev.\  D {\bf 82} (2010) 086010
  [arXiv:1005.5134 [hep-ph]].
  %%CITATION = PHRVA,D82,086010;%%


\bibitem{CGQ3} 
   J.~A.~Cabrer, G.~von Gersdorff and M.~Quiros,
  %``Suppressing Electroweak Precision Observables in 5D Warped Models,''
  arXiv:1103.1388 [hep-ph].
  %%CITATION = ARXIV:1103.1388;%%

%\cite{Peskin:1991sw}
\bibitem{Peskin:1991sw}
  M.~E.~Peskin and T.~Takeuchi,
  %``Estimation of oblique electroweak corrections,''
  Phys.\ Rev.\  D {\bf 46} (1992) 381.
  %%CITATION = PHRVA,D46,381;%%

\bibitem{Nakamura:2010zzi}
   K.~Nakamura  [Particle Data Group],
  %``Review of particle physics,''
  J.\ Phys.\ G {\bf 37} (2010) 075021.
  %%CITATION = JPHGB,G37,075021;%%

%\cite{Round:2010kj}
\bibitem{Round:2010kj}
 M.~Round,
  %``Holographic Renormalisation and the Electroweak Precision Parameters,''
  Phys.\ Rev.\  D {\bf 82} (2010) 053002
  [arXiv:1003.2933 [hep-ph]].
  %%CITATION = PHRVA,D82,053002;%%

\bibitem{Klebanov:2000nc}
  I.~R.~Klebanov and A.~A.~Tseytlin,
  %``Gravity Duals of Supersymmetric SU(N) x SU(N+M) Gauge Theories,''
  Nucl.\ Phys.\  B {\bf 578} (2000) 123
  [arXiv:hep-th/0002159].
  %%CITATION = NUPHA,B578,123;%%

\bibitem{Brummer:2005sh}
  F.~Brummer, A.~Hebecker and E.~Trincherini,
  %``The throat as a Randall-Sundrum model with Goldberger-Wise
  %stabilization,''
  Nucl.\ Phys.\  B {\bf 738} (2006) 283
  [arXiv:hep-th/0510113];
  %%CITATION = NUPHA,B738,283;%%
%
%\bibitem{Hassanain:2007js}
  B.~Hassanain, J.~March-Russell and M.~Schvellinger,
  %``Warped Deformed Throats have Faster (Electroweak) Phase Transitions,''
  JHEP {\bf 0710} (2007) 089
  [arXiv:0708.2060 [hep-th]].
  %%CITATION = JHEPA,0710,089;%%


\bibitem{Luty:2004ye}
  M.~A.~Luty and T.~Okui,
  %``Conformal technicolor,''
  JHEP {\bf 0609} (2006) 070
  [arXiv:hep-ph/0409274].
  %%CITATION = JHEPA,0609,070;%%


\bibitem{DeWolfe:1999cp}
  O.~DeWolfe, D.~Z.~Freedman, S.~S.~Gubser and A.~Karch,
  %``Modeling the fifth dimension with scalars and gravity,''
  Phys.\ Rev.\  D {\bf 62} (2000) 046008
  [arXiv:hep-th/9909134].
  %%CITATION = PHRVA,D62,046008;%%


\end{thebibliography}
\end{document}